\documentclass[square,aip,preprint,showkeys,superscriptaddress]{revtex4}
\usepackage{graphicx}
\usepackage{epstopdf,amsmath}

\begin{document}
	\title{YBCO-based non-volatile ReRAM tested in Low Earth Orbit}
	\author{C. Acha}
	\thanks{corresponding author (acha@df.uba.ar)}
	\affiliation{Laboratorio de Bajas Temperaturas, Departamento de
		F\'{\i}sica, FCEyN, Universidad de Buenos Aires and IFIBA,
		UBA-CONICET, Pabell\'on I, Ciudad Universitaria, C1428EHA CABA, Argentina}
	\author{M. Barella}
	\thanks{Present address: Centro de Investigaciones en Bionanociencias (CIBION), Consejo Nacional de Investigaciones Cient\'{\i}ficas y T\'ecnicas (CONICET), Godoy Cruz 2390, C1425FQD CABA, Argentina}
	\affiliation{Escuela de Ciencia y Tecnolog\'{\i}a, Universidad Nacional de San Mart\'{\i}n,
		Mart\'{\i}n de Irigoyen 3100, B1650JKA San Mart\'{\i}n, Bs. As.,
		Argentina} \affiliation{Consejo Nacional de Investigaciones
		Cient\'{\i}ficas y T\'ecnicas (CONICET), Godoy Cruz 2290, C1425FQB
		CABA, Argentina}
	\author{G. A. Sanca}
	\affiliation{Escuela de Ciencia y Tecnolog\'{\i}a, Universidad Nacional de San Mart\'{\i}n,
		Mart\'{\i}n de Irigoyen 3100, B1650JKA San Mart\'{\i}n, Bs. As.,
		Argentina}
	\author{F. Gomez Marlasca}
	\affiliation{Comisi\'on Nacional de Energ\'{\i}a At\'omica (CNEA),
		Av. Del Libertador 8250, C1429BNP CABA, Argentina}
	\author{H. Huhtinen}
	\affiliation{Wihuri Physical Laboratory, Department of Physics and Astronomy,
		University of Turku, FI-20014 Turku, Finland}
	\author{P. Paturi}
	\affiliation{Wihuri Physical Laboratory, Department of Physics and Astronomy,
		University of Turku, FI-20014 Turku, Finland}
	\author{P. Levy}
	 \affiliation{Escuela de Ciencia y Tecnolog\'{\i}a, Universidad Nacional de San Mart\'{\i}n,
		Mart\'{\i}n de Irigoyen 3100, B1650JKA San Mart\'{\i}n, Bs. As.,
		Argentina}
	\affiliation{Consejo
		Nacional de Investigaciones Cient\'{\i}ficas y T\'ecnicas (CONICET),
		Godoy Cruz 2290, C1425FQB CABA, Argentina}
	\affiliation{Comisi\'on Nacional de Energ\'{\i}a At\'omica (CNEA),
		Av. Del Libertador 8250, C1429BNP CABA, Argentina} 
	\author{F. Golmar}
	\affiliation{Escuela de Ciencia y Tecnolog\'{\i}a, Universidad Nacional de San Mart\'{\i}n,
		Mart\'{\i}n de Irigoyen 3100, B1650JKA San Mart\'{\i}n, Bs. As.,
		Argentina} \affiliation{Consejo Nacional de Investigaciones
		Cient\'{\i}ficas y T\'ecnicas (CONICET), Godoy Cruz 2290, C1425FQB
		CABA, Argentina}
	
	\date{\today}

\begin{abstract}
	
	An YBCO-based test structure corresponding to the family of ReRAM devices associated with the valence change mechanism is presented. We have characterized its electrical response previous to its lift-of{}f to a Low Earth Orbit (LEO) using standard electronics and also with the dedicated LabOSat-01 controller. Similar results were obtained in both cases. \\
	After about 200 days at LEO on board a small satellite, electrical tests started on the memory device using the LabOSat-01 controller. We discuss the results of the first 150 tests, performed along a 433-day time interval in space. The memory device remained operational despite the hostile conditions that involved launching, lift-of{}f vibrations, permanent thermal cycling and exposure to ionizing radiation, with doses 3 orders of magnitude greater than the usual ones on Earth. The device showed resistive switching and IV characteristics similar to those measured on Earth, although with changes that follow a smooth drift in time. A detailed study of the electrical transport mechanisms, based on previous models that indicate the existence of various conducting  mechanisms through the metal-YBCO interface showed that the observed drift can be associated with a local temperature drift at the LabOSat controller, with no clear evidence that allows determining changes in the underlying microscopic factors. These results show the reliability of complex-oxide non-volatile ReRAM-based devices in order to operate under all the hostile conditions encountered in space-borne applications.

\end{abstract}


\maketitle

\section{INTRODUCTION}
Many efforts have been made in recent years to deepen the study of
the properties of memory devices based on the resistance switching
effect (called ReRAM or Memristors), in order to analyze their
possible applications as memory devices as well as in logic
circuits~\cite{Borghetti10} or in those that mimic the electrical
behavior of synapses~\cite{Jo10} or even of neurons~\cite{Stoliar17}.
Another area of great technological importance is related to the
development of radiation resistant memories~\cite{Dodd10}. This is of particular interest for aerospace industry, as electronic circuits in space must be protected from ionizing radiation from the Sun and/or from other radiation sources. Depending on spacecraft's orbit, different strategies are employed to mitigate radiation effects~\cite{Cardarilli03,Gerardin10}. For instance, in Low Earth Orbit (LEO) missions, long-term radiation effects, like Total Ionizing Dose (TID), is one of the main concerns. Ionizing radiation in LEO is mainly composed of high fluxes of energetic protons and electrons, trapped in the inner and outer Van Allen belts. Typically, most of this ionizing particles are stopped by aluminum shields, albeit this strategy increases spacecraft's weight and, in consequence, mission's cost. In the case of interplanetary or deep-space missions, main risks are transient events, like Single Event Effects, caused by high energy particles from the Sun (coronal mass ejection) or galactic cosmic rays. Usually, this missions use fail-safe systems that rely on radiation-hardened electronics, error correction algorithms and redundant CMOS circuits to mitigate this kind of effects.

Ubiquitous flash memory technology for non volatile storage relies on charge confinement, which is unstable against ionizing radiation~\cite{Gerardin13}. Shielding, redundancy and watchdog timers are common strategies used to mitigate sporadic but profoundly disturbing problems triggered by the incidence of foreign radiation. At the core of the alternative ReRAM technology, memristors offer either interface or filamentary based mechanisms that rely on their constituent materials properties, thus exhibiting hardened response against ionizing radiation~\cite{Velo17}.  Moreover, due to their simple and downsizeable capacitor like geometry, other potential problems are expected to be tackled: robustness against launch vibrations, and stability against omnipresent thermal cycling.

The resistance switching (RS) of ReRAM devices based on perovkite oxide-metal interfaces is associated with the local valence change induced by the electric field-driven migration of oxygen vacancies~\cite{Waser09a}. Their electrical conduction is mainly determined by the microscopic properties of the volume of the oxide, close to the interface~\cite{Fu14,Rosillo17}, rather than to the interfacial surface itself, as occurs for example in cases where a Schottky barrier is formed~\cite{Jeong09}. This fact can be exploited to produce memory devices resistant to radiation considering that it will be more difficult for radiation to affect a volume property than one that depends solely on the surface. With this strategy in mind,  in this work we explore the electrical response of YBa$_2$Cu$_3$O$_{7-\delta}$ (YBCO)-based memristors at a Low Earth Orbit (LEO). The electrical tests were performed by a LabOSat-01 controller (shortly LS01 hereafter)~\cite{Barella16} at a 500 km-height orbit onboard a small satellite of the Satellogic company~\cite{Satellogic}. The electronic controller is powered once a day: a Standard Test (ST) is carried out and the results are stored for transmission to Earth through the satellite's communication capabilities. There is no synchronicity between this measurement and the position of the satellite, thus our results are to be analyzed as performed at random sampling at a unique LEO. 

Our results indicate that the YBCO-based test structure remained operational after more than 14 months in LEO, showing similar characteristics to those previously measured on Earth. A detailed analysis of the electrical transport, based on previous models, allowed us to infer that the small drift over time observed in the electrical properties can be essentially associated with the temperature changes measured onboard the satellite at LEO and that there is no evidence of microscopic changes such as those related to the increase of defects, both by thermal cycling or ionizing radiation, that could affect the performance of the device. In that sense, our studies serve as an initial step towards the validation of perovskite oxide-based ReRAM devices for space-borne applications.

\section{EXPERIMENTAL DETAILS}

Fully relaxed YBCO thin films were grown by pulsed laser deposition
(PLD) on top of a (100) single crystal STO substrate. The deposition
was performed by applying 1500 pulses with a growth rate of 0.1
nm/pulse, producing a 150 nm YBCO thick layer. This growth rate was previously confirmed by transmission electron microscopy (TEM), under the same deposition conditions~\cite{Khan19}. The
superconducting transition temperature ($\sim$ 90 K) was determined
by resistivity and magnetization measurements, confirming that the
YBCO films are near optimally-doped. Additional details of their
synthesis and characterization can be found
elsewhere~\cite{Huhtinen01,Paturi04,Peurla07,Khan19}.

\begin{figure}
	\vspace{0mm}
	\centerline{\includegraphics[angle=0,scale=0.4]{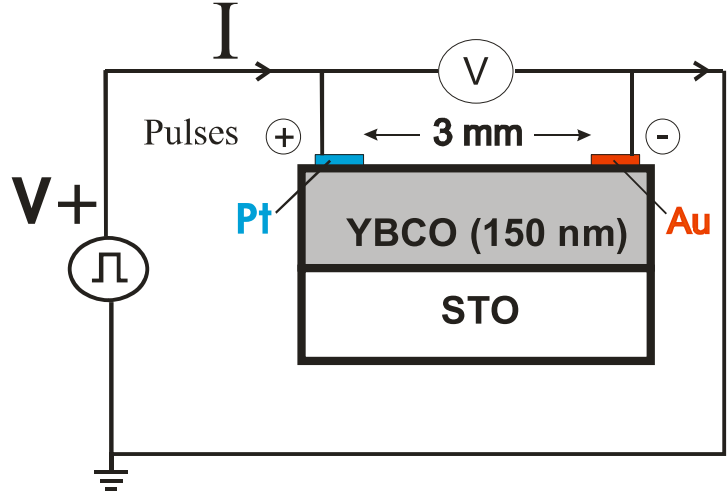}}
	\vspace{-10mm}\caption{(Color online) Circuit used to apply writing and reading pulses and scheme of the YBCO thin film-based device with its electrodes (not in scale).} \vspace{-0mm}
	\label{fig:devices}
\end{figure}

The device under test (DUT) was built by sputtering 30 nm of 2 different metal contacts (0.7 x 0.7 mm$^2$) on top of one YBCO film surface, arranged in a planar structure (which maximizes the exposure to external irradiation), as depicted in Fig.~\ref{fig:devices}.  One of the contacts, labeled arbitrarily as ``+", was made with Pt, while Au was used for the ground (-) pad. We have used Pt and Au in order to produce a DUT with essentially only one active interface, as will be described later. Electrodes have a mean separation of 3 mm.  Cu leads were carefully fixed over them by using silver paint without contacting directly the surface of the YBCO sample. The YBCO-based DUT was attached inside a SOIC-16 package, where its Cu leads were bonded with conductive silver paint. Finally, the package was sealed using space-qualified epoxy resin~\cite{Barella18}.

\begin{figure} [hbt]
	\vspace{-10mm}
	\centerline{\includegraphics[angle=0,scale=0.5]{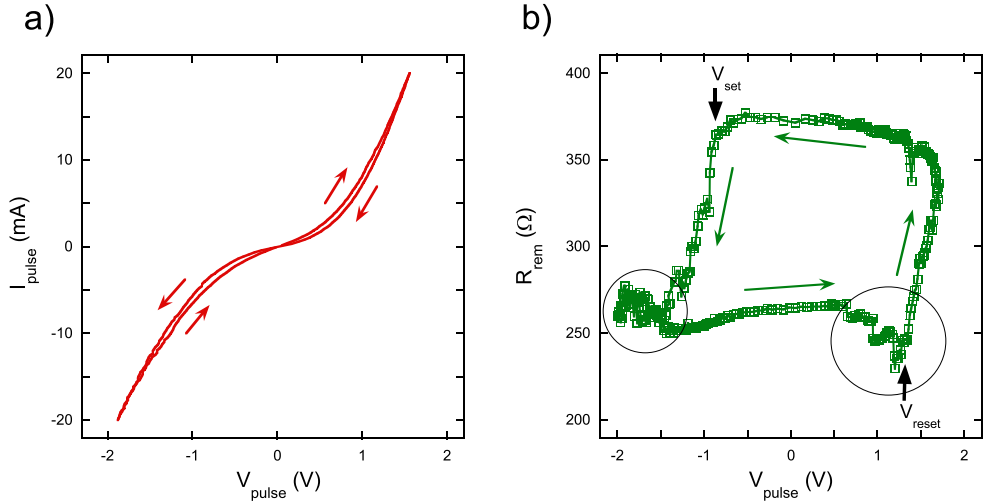}}
	\vspace{-20mm}\caption{(Color online) a) IV characteristics and b) RHSL of DUT  measured previous to the lift-off, showing the existence of non-linear effects, hysteresis and consequently, resistive memory. The colored arrows indicate the circulation path. The counter-clock-wise circulation in b) indicates that most of the switching occurs at the Au/YBCO (-) interface. Circles indicate a minor loop producing the shape called ``Table with legs", associated with the complementary RS of the opposite interface. The voltages V$_{set}$ and V$_{reset}$ are also indicated. }
	\vspace{-0mm}
	\label{fig:RHSL_LBT1}
\end{figure}

The DUT was first characterized at room
temperature at the Laboratorio de Bajas Temperaturas (LBT) by using a B2902B Agilent SMU, programmed to apply 10 ms
current writing pulses (I$_{pulse}$) of increasing and decreasing
amplitude between $\pm$ 20 mA , establishing a hysteresis cycle by measuring the voltage during the pulses (V$_{pulse}$). In this way current-voltage (IV)
characteristics of the DUT are measured. With 1 s delay after
each writing voltage pulse, a small reading voltage is applied and
again, by measuring the current, the remanent resistance (R$_{rem}$)
is determined. By plotting R$_{rem}$ as a function of V$_{pulse}$ a
resistance hysteresis switching loop (RHSL) can be observed, with V$_{set}$ and V$_{reset}$ as the voltages where the resistance switching begins, producing a low resistance or a high resistance state, respectively, as
shown in Fig.~\ref{fig:RHSL_LBT1}. As obtained previously for ceramics~\cite{Schulman13} and for thin
films~\cite{Lanosa20} metal/YBCO devices, our DUT exhibit a
bipolar RS and hysteretical and non-linear IV
characteristics. The counter-clockwise RHSL indicates that the
``active" contact (ie the one that generates most of the resistance
change) is, as expected, the ground Au contact, although a small change of
the proper ``+" device contacts is also observed (marked with circles
in Fig.~\ref{fig:RHSL_LBT1}), forming the shape coined as ``table with
legs"~\cite{Strukov08}.

To characterize the packaged-DUT in orbit, we soldered it onto a dedicated controller, the LS01 board~\cite{Barella16}. This board was specifically designed to electrically test two and three terminal electronic devices in hostile environments. Its purpose is to increase the Technology Readiness Level (TRL) of electronic devices for space-borne applications. In fact, LS01 has proven to be successful characterizing TiO$_{2}$-based and La$_{1/3}$Ca$_{2/3}$MnO$_{3}$-based ReRAMs in LEO \cite{Barella19}. It uses a SMU to test the DUT and has several sensors to monitor the hostile conditions while it is operating \cite{Barella16}. Particularly, in this work, we report data from its temperature sensor and 3 solid-state dosimeters, which measure long-term radiation doses, i.e., TID using COTS pMOS transistors~\cite{Lipo07,Inza09,Sanca17}. In short, when a CMOS transistor is exposed to ionizing radiation, its threshold voltage (V$_{th}$) shifts to negative values due to charge accumulation in oxide traps under the gate structure \cite{Oldham03}. \\

Let us now consider how LS01 performs the measurements on the DUT, while in orbit. At most once a day, the host satellite triggers a Standard Test (ST) in LS01. During a ST several experiments are run. Among them, temperature and accumulated radiation dose are registered, and IV and RHSL measurements are performed on the DUT. Although the aim was to replicate the measurements made on the DUT at LBT as accurately as possible, limitations related to the measurement platform arose. Namely, (i) the width of the current writing pulses $I_{pulse}$ is larger (100 ms instead of 10 ms). (ii) Only one hysteretic IV and one RHSL curve are performed per ST, as opposed to what happens at LBT where many of these curves are measured successively. This means that the relaxation-related features that emerge when considering a set of successive curves measured in orbit (sent in different STs) might look different from the features that emerge in a set of curves measured at the lab. Finally (iii), the number of writing pulse values downloaded to Earth---along with the $(V_{pulse}, I_{pulse})$ and $(V_{pulse}, R_{rem})$ values needed to plot IV and RHSL curves---is smaller than the actual number of writing pulses $I_{pulse}$ applied to the DUT in orbit. The reason for (ii) and (iii) is that the amount of data LS01 is able to send to Earth is limited, so the available amount of data storage has to be divided among the different experiments taking place onboard. Regarding (iii), the reason the number of pulses \textit{applied} and the number of pulses \textit{recorded} are different, is that observing similar hysteretic behavior in different experimental setups requires replicating the same pulsing sequence in both setups. Even when LS01 is technically capable of replicating the same pulsing sequence, the amount of data generated by it might become too large to be downloaded from space. In the experiments we are about to discuss, data related to roughly 1 out of 4 pulses is recorded and sent to Earth. The pulsing sequence to measure IV and RHSL curves embedded in LS01 starts by applying a +250 $\mu$A pulse,  then the amplitude of the pulses  is increased in steps of  250 $\mu$A  until it reaches a maximum of 20 mA; after that, the amplitude is decreased back to  +250 $\mu$A, also in 250 $\mu$A steps. Once the positive part of the sequence is swept, an analog procedure is performed for the negative part, until a minimum of -17 mA is reached in steps of -250 $\mu$A. 250 ms after each writing pulse is applied, a reading pulse of $I_{rem} = 1$ mA is applied.  Simultaneous measurements of $V_{pulse}$  ($V_{rem}$) are taken while applying every $I_{pulse}$ ($I_{rem}$) pulse; these measurements are used to determine the IV and RHSL curves. 

\begin{figure} [hbt]
	\vspace{5mm}
	\centerline{\includegraphics[angle=0,scale=0.45]{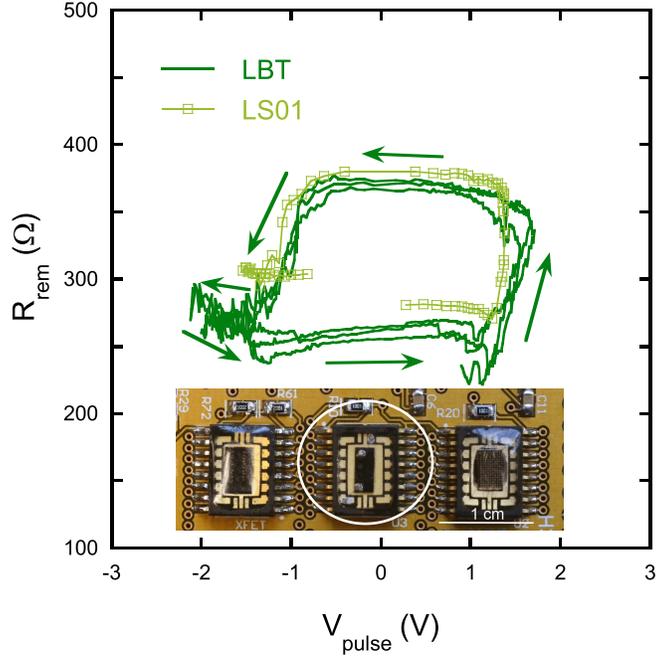}}
	\vspace{-4mm}\caption{(Color online) Comparison of the first RHSLs measured at the LBT and by LS01 on Earth at room temperature ($\sim 25 ^\circ$C, I$_{rem} =$ 1 mA), before lift-off. The inset shows a detail of the LS01 board, where the packaged YBCO-based device is located with other DUTs (center).} \vspace{-0mm}
	\label{fig:1eros}
\end{figure}

Fig.~\ref{fig:1eros} depicts RHSL data acquired using the LS01 board before lift-off and using the aforementioned commercial electronics at the LBT. The inset of this figure shows the packaged YBCO-based DUT (at the center) soldered onto the board. Despite the typical variability observed in successive RHSLs, both sets of data are similar, making evident the actual sensitivity of LS01.

Next, LS01 was integrated into NuSat-5 satellite (COSPAR ID: 2018-015K) of the Satellogic company ~\cite{Satellogic}. The rocket carrying the satellite lifted-off from Jiuquan Space Center (China) on February 5th, 2018 (day number 0 of the experiment). Once in orbit, LS01 stayed 210 days on stand-by. The first experiment over the DUT was executed on August 31st, 2018. Depending on mission task priorities, not all the ST are downloaded from space. In this way, 148 experiments executed over the DUT were downloaded, up to April 11th, 2019. This comprises a total of 433 days in orbit ($\sim$ 1.2 yr).

\begin{figure} [hbt]
	\vspace{-5mm}
	\centerline{\includegraphics[angle=0,scale=0.4]{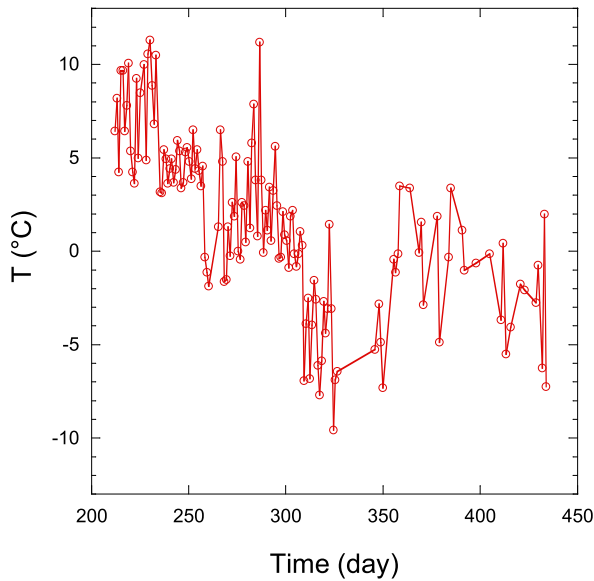}}
	\vspace{-2mm}\caption{(Color online) Temperature evolution of LS01 board  as a function of the number of days at LEO. Data is not regularly distributed as not all the ST are downloaded from space.} \vspace{-0mm}
	\label{fig:Tvsday}
\end{figure}

As days in orbit went by, we observed LS01's temperature ranging mostly between -10 to 13 $^\circ$C as can be seen in Fig.~\ref{fig:Tvsday}. Although periodic variations are expected, probably related to orbit's cinematic ($\sim$ 90 min/orbit) and to the 24 h ST execution time lapse, a smooth shift toward lower temperatures is observed as the number of days at LEO increases. Fig.~\ref{fig:Tvsday} is a partial sample of the overall thermal cycling that is reproduced with each orbit, that may not display the maximums and minimums reached, in some cases due to repositioning of the satellite while performing other tasks. 
This gives rise to additional stress on the DUT and may also affect all the associated measurement electronics. \\

During this period, LS01 dosimeters did not show significant threshold voltage shifts. The standard deviation of each dosimeter's dataset was comparable to 1 Gy(Si). Nevertheless, we simulated TID using SPENVIS tool \cite{Spenvis04} to estimate expected radiation levels at the satellite orbit ($\sim$ 500 km altitude). 

\begin{figure} [h]
	\vspace{2mm}
	\centerline{\includegraphics[angle=0,scale=0.35]{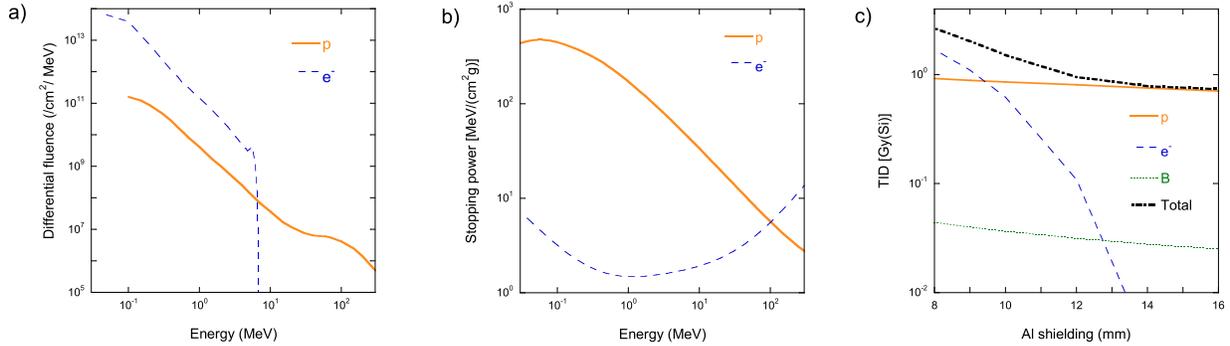}}
	\vspace{-2mm}\caption{(Color online) a) Differential fluence as a function of the energy of protons (p) and electrons (e$^-$).  b) Aluminium stopping power for different particles as a function of their energy. c) Simulated Total Ionizing Dose (TID) for the mission’s period, as a function of Al shielding provided by the satellite structure ($\sim$9 mm). Different sources contribute to TID, i.e. mainly electrons, protons and Bremsstrahlung radiation (B). Solar protons fluence was neglible during this period.} \vspace{-0mm}
	\label{fig:rad}
\end{figure}

To perform the simulations, we used the exact dates reported 
above and the orbital parameters of the NuSat-5 satellite (a 500.1 km altitude circular orbit, with 97.33$^{\circ}$ inclination, 11.21$^{\circ}$ right ascension of ascending node, 251.6$^{\circ}$ argument of perigee, and a true anomaly of 133.3$^{\circ}$). Fluence and doses were simulated using an effective shielding equivalent to 9 mm-thick aluminum foam panes, AP-8 and AE-8 models for trapped protons and electrons, and ESP-PSYCHIC model for solar particles. We also consider minimum solar activity (end of Solar cycle 24), and for TID calculations we used SHIELDOSE-2 model. \\
The simulated differential fluences of ionizing particles found in the orbit of the satellite are shown in Fig.~\ref{fig:rad}a for the total mission's period. For energies below 10 MeV, trapped electron fluence is around 2 orders of magnitude higher than proton's. Conversely, for higher energies, electron fluence is significantly reduced which causes that high-energy contribution to TID come mainly from trapped protons. However, this contrast in the trapped particle fluence spectra is balanced by the thick aluminum shielding and its stopping power (see Fig.~\ref{fig:rad}b). As can be seen in Fig.~\ref{fig:rad}c, the total TID experienced by the DUT inside the satellite is composed of similar fractions of proton and electron TID.

Finally, the simulations indicate that the absorbed dose should be around 2 Gy(Si) for the total mission's period. This result is slightly lower than expected values for a 1-yr period of typical LEO missions (see references \cite{Lipo07,Barella19} and references therein). However, as Huston and Pfitzer pointed out~\cite{Huston13}, we should consider this result as a rough approximation, as the trapped particle models used here returned overestimated predictions up to a factor of 2 in those previous works. Hence, it is not surprising that LS01 dosimeters did not sense critical levels of TID.

\section{RESULTS AND DISCUSSION}

\begin{figure} [h]
	\vspace{0mm}
	\centerline{\includegraphics[angle=0,scale=0.4]{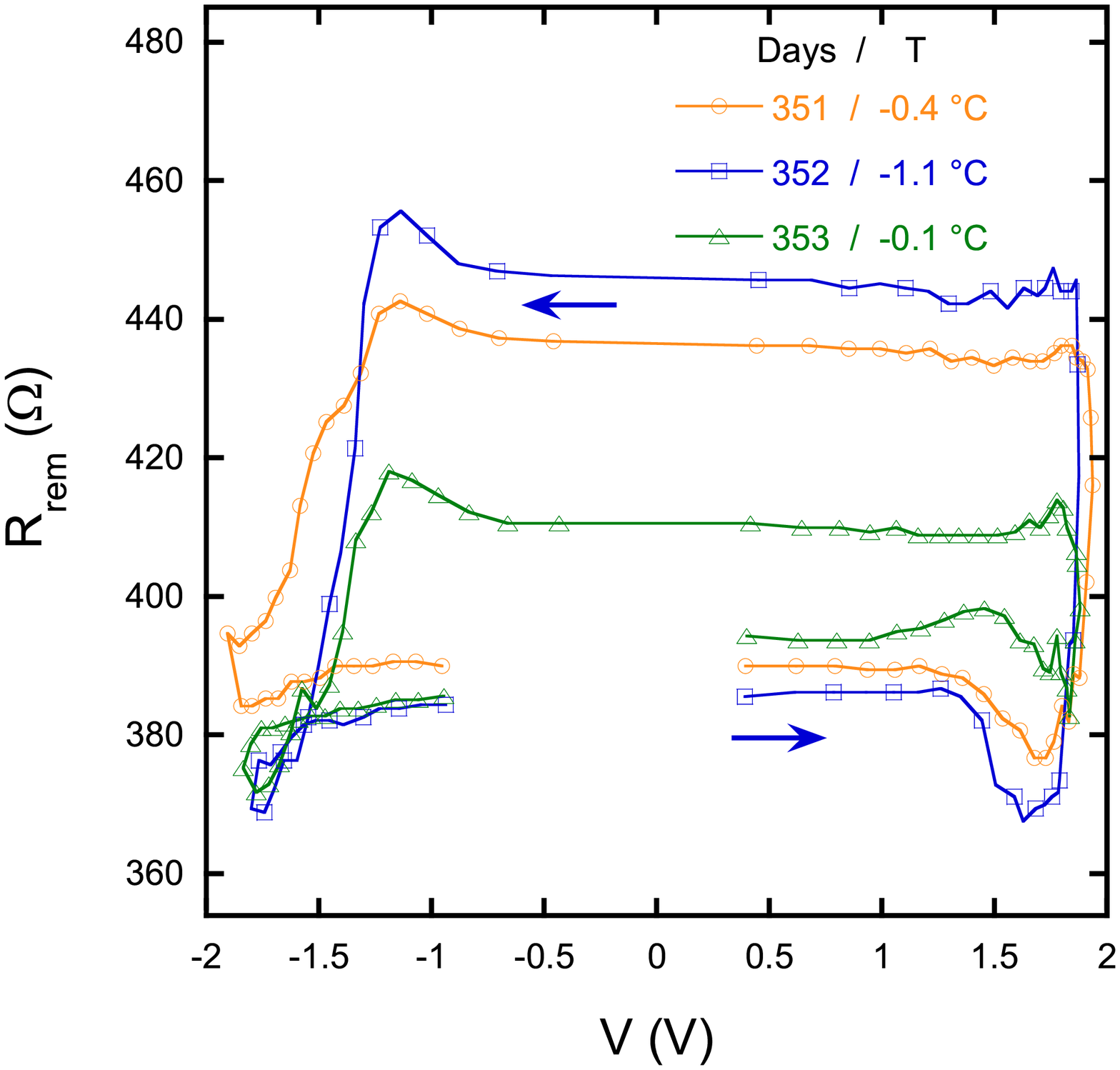}}
	\vspace{-3mm}\caption{(Color online) Typical succesive RHSLs of DUT measured after almost 1 year at LEO, showing similar features as those observed before lift-off, such as the counter-clockwise circulation and the table with legs structure.} \vspace{-0mm}
	\label{fig:Rrem_vs_day}
\end{figure}

Typical successive RHSLs measured by LS01 after nearly 1 year at LEO are shown in Fig.~\ref{fig:Rrem_vs_day}. The DUT still shows bipolar resistive switching and the counter-clockwise circulation is maintained as well as the shape ``Table with Legs", indicating that the Au/YBCO interface is still the dominant with a lower resistive switching contribution of the Pt/YBCO interface. A small change in the remanent resistance can be observed after each cycle. This can be a consequence of the temperature variation of the DUT or can be related to a relaxation of the final resistance, considering the 24 h delay between successive measurements. This effect is characteristic of the YBCO-based interfaces, indicating the high mobility of oxygens along specific crystallographic orientations or in grain boundaries~\cite{Schulman12,Placenik12}. Additionally, it can be noticed that the remanent resistance values are 15-30\% higher than those measured before lift-off. \\

\begin{figure} [htb]
	\vspace{-5mm}
	\centerline{\includegraphics[angle=0,scale=0.4]{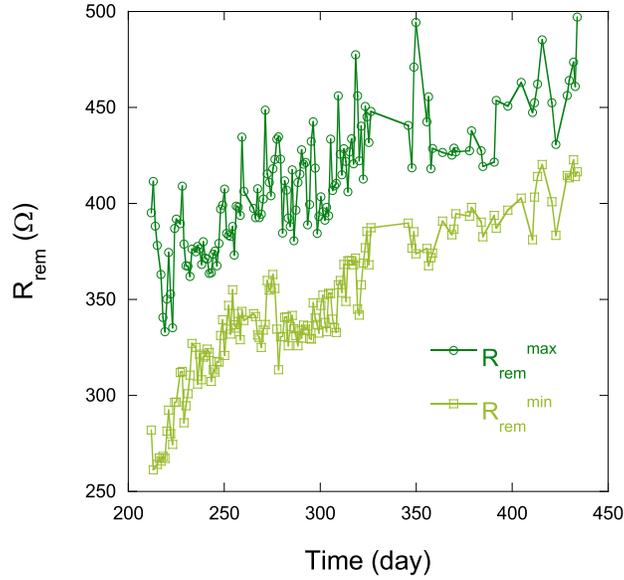}}
	\vspace{-3mm}\caption{(Color online) Maximum and minimum values of the remanent resistance (R$_{rem}$) extracted from all the RHSLs as a function of the days at LEO. Apart from the large fluctuations observed both magnitudes show a tendency to increase with increasing number of days.} \vspace{-0mm}
	\label{fig:Rremmaxmin_vs_day}
\end{figure}

\begin{figure} [htb]
	\vspace{0mm}
	\centerline{\includegraphics[angle=0,scale=0.4]{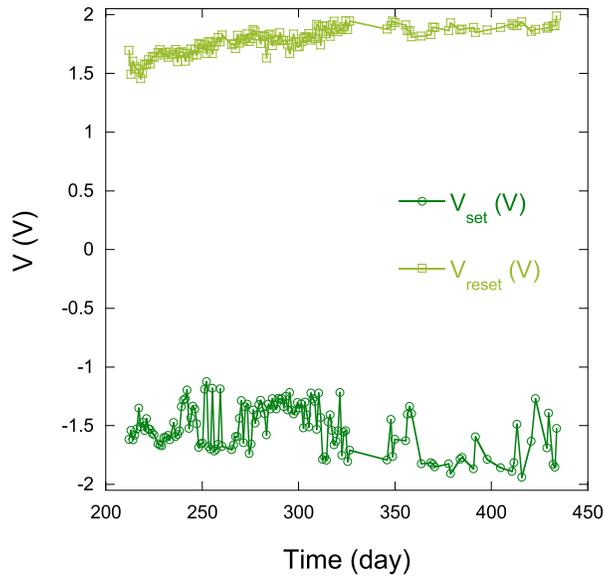}}
	\vspace{-3mm}\caption{(Color online) Set and reset voltages extracted from the RHSL  as a function of the days at LEO. A tendency to increase in absolute value with the number of days is observed.} \vspace{-0mm}
	\label{fig:Vset_reset_vs_day}
\end{figure}

In order to gain insight on the origin of the variations observed in the RHSLs, we plotted in Fig.~\ref{fig:Rremmaxmin_vs_day} the evolution of the maximum and minimum R$_{rem}$ as a function of the days at LEO. A noisy behavior with an overall tendency to increase with increasing the number of days can be observed. A similar tendency can be observed in Fig.~\ref{fig:Vset_reset_vs_day} for the voltages V$_{set}$ and V$_{reset}$, although the variation of V$_{set}$ with the number of days at LEO is less evident due to its noisy behavior.

\begin{figure} [htb]
	\vspace{0mm}
	\centerline{\includegraphics[angle=0,scale=0.4]{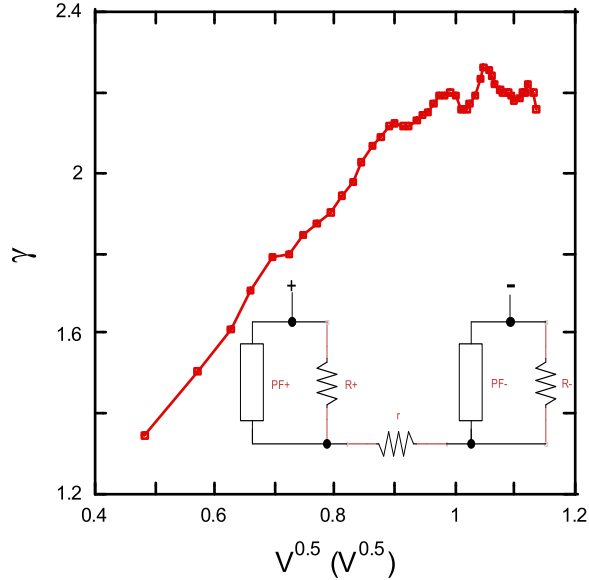}}
	\vspace{-3mm}\caption{Power exponent parameter $\gamma$ as a function of the square root of the voltage amplitude of the pulses. The inset shows the equivalent circuit model for the DUT. Each contact ($\pm$) can be represented by a Poole-Frenkel element in parallel with a leakage resistor (R), both in series with a smaller resistor (r). } \vspace{-0mm}
	\label{fig:gamma_equiv}
\end{figure}


In order to understand the physical origin of the observed evolution of both characteristic parameters of this memristive DUT, we can deepen our analysis trying to determine the conduction mechanisms involved in each interface. For this, we can appeal to the analysis of the IV characteristic curves based on the power exponent   $\gamma=dLn(I)/dLn(V)$ plotted as a function of
$V^{1/2}$~\cite{Acha17}. Indeed, this method has been very helpful in determining the existence of different transport mechanisms present in a junction, especially when more than one is involved~\cite{Acha16,Acevedo17,Ghenzi19}.

The $\gamma$ representation of our DUT, determined before to lift-off, is shown in Fig.~\ref{fig:gamma_equiv}. The almost linear dependence from low voltages (with a positive intercept) and up to a voltage where a maximum is reached, confirms that a Poole-Frenkel (PF) emission is the main conduction mechanism through the interfaces of our DUT. A more detailed equivalent circuit model, which includes a leak resistor (R$^{\pm}$) in parallel with the non-linear PF$^{\pm}$ element and a series resistor (r), representing the interface-bulk resistance plus the intrinsic resistance of the film, is presented in the inset of Fig.~\ref{fig:gamma_equiv}.  $``+"$ and $``-"$ represent the Pt/YBCO and the Au/YBCO interfaces, respectively. This more elaborated model was determined in previous studies performed on the very same interfaces~\cite{Acha11,Schulman15,Lanosa20}.

Within this framework, the current through each PF element ($I_{PF}^{\pm}$) as a function of the voltage $V_{PF}^{\pm}$ at a fixed temperature $T$, can be expressed as~\cite{Simmons67}:

\begin{equation}
\label{eq:PF} I_{PF}^{\pm} = (R_0^{\pm})^{-1} V_{PF}^{\pm} \exp[-E_{Trap}^{\pm}/(k_B T) + B^{\pm}
(V_{PF}^{\pm})^{1/2}],
\end{equation}
\noindent with
\begin{equation}
B^{\pm} = \frac{q^{3/2}}{k_B T (\pi \epsilon'^{\pm} d^{\pm})^{1/2}},
\end{equation}

\noindent where $R_0^{\pm}$ is a pre-factor associated with the geometric factor of the conducting path, the
electronic drift mobility ($\mu$) and the density of states in the
conduction band. $E_{Trap}^{\pm}$ is the trap energy level, $k_B$ the Boltzmann
constant, $q$ the electron charge, $\epsilon'^{\pm}$ the real part of the dielectric constant of the oxide and $d^{\pm}$ the interfacial thickness where most of the voltage drops (for each interface ${\pm}$). In this way, the voltage-dependent resistance related to the PF element of each interface can then be expressed as:

\begin{equation}
\label{eq:PF2} R_{PF}^{\pm} = \frac{V_{PF}^{\pm}}{I_{PF}^{\pm}} = R_0^{\pm} \exp[E_{Trap}^{\pm}/(k_B T) - B^{\pm}
(V^{\pm})^{1/2}] \overset{V\rightarrow 0}\simeq R_0^{\pm} \exp[E_{Trap}^{\pm}/(k_B T)].
\end{equation}

\begin{figure} [htb]
	\vspace{-18mm}
	\centerline{\includegraphics[angle=0,scale=0.48]{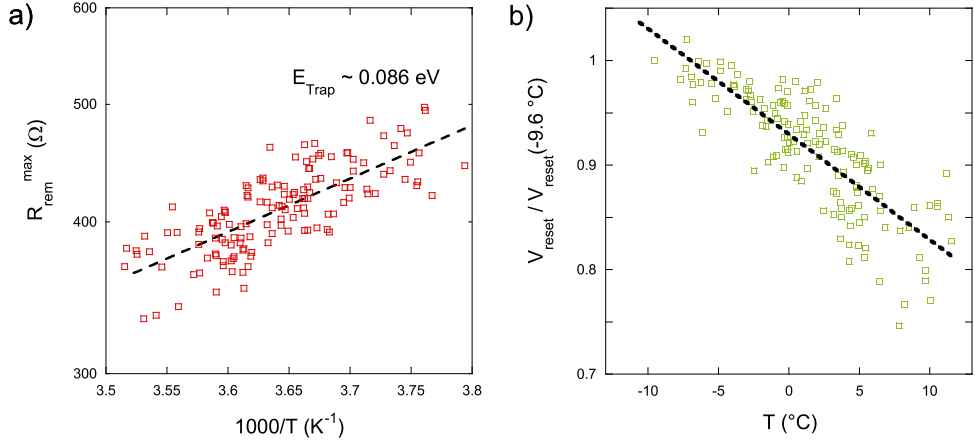}}
	\vspace{-18mm}\caption{(Color online) a) Maximum remanent resistance R$_{rem}^{max}$ (log scale) as a function of T$^{-1}$, where $T$ is the LS01 board temperature during DUT measurement. The dashed line is an exponential fit which represent the temperature dependence of the Poole-Frenkel element (see Eq.~\ref{eq:PF2}). The slope indicates the mean energy of the electrical carrier traps (E$_{Trap}$). b) Normalized reset voltages of both devices plotted as a function of $T$.} \vspace{-0mm}
	\label{fig:Rrem_Vreset_vs_T}
\end{figure}

This equation indicates that in the low-voltage limit, the PF element behaves, as a function of temperature, as a semiconductor does. Despite the existence of the two interfaces and the parallel and serial resistances indicated in the more detailed circuit model, if we plot the measured R$_{rem}^{max}$ as a function of the temperature of each day at LEO (see Fig.~\ref{fig:Rrem_Vreset_vs_T}a), we can observe that our DUT behaves as indicated by Eq.~\ref{eq:PF2}. In fact, this result is also indicating that the resistive behavior of the DUT is dominated by the Au/YBCO interface and more particularly by the PF emission linked to the oxide in the interfacial zone close to Au. The low resistance of both the Pt/YBCO interface (probably associated with a low R$^+$ value) and the film intrinsic resistance (r) as well as the low ohmic conducting leakage through the  R$^-$ element determine the simplicity of the obtained dependence.

The obtained value for E$_{Trap} \simeq 0.086~ eV$ is in close accordance with the values already obtained for the Au/YBCO interface in its high resistance state~\cite{Schulman15}. Besides the noisy behavior of the resistance of the DUT along the days in the hostile environment at LEO, the observed drift of the remanent resistance seems to be strongly associated with the temperature variations that LS01 experiences within the satellite. This reasoning can be applied qualitatively to the high resistance state of the successive RHSLs presented in Fig.~\ref{fig:Rrem_vs_day}, where the value of R$_{rem}^{max}$ can be ordered inversely to the temperature at which the measurement was made. In a similar way, if we plot V$_{reset}$ as a function of the temperature measured by LS01, the data seems to follow an almost linear dependence with a negative slope, as shown in Fig.~\ref{fig:Rrem_Vreset_vs_T}b. This dependence was already observed and previously reported~\cite{Acha09b}.

\begin{figure} [htb]
	\vspace{0mm}
	\centerline{\includegraphics[angle=0,scale=0.4]{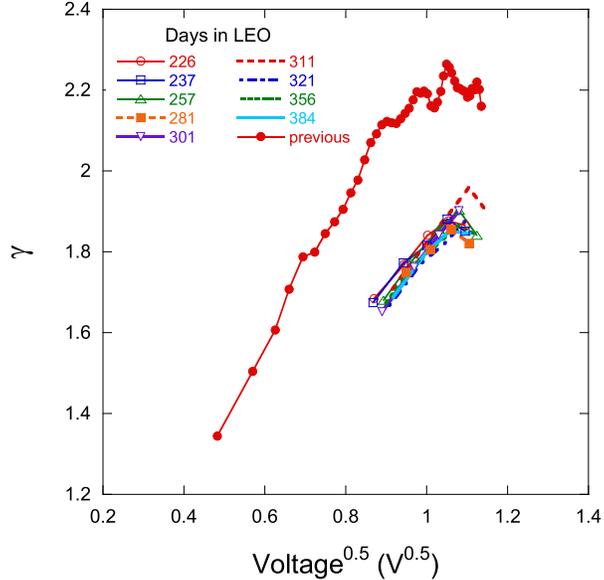}}
	\vspace{-2mm}\caption{(Color online) Comparison of the $\gamma$ parameter as a function of
		V$^{1/2}$ for the DUT measured previously at LBT and by LS01 as a function of the days at LEO.} \vspace{-0mm}
	\label{fig:gamma_compara}
\end{figure}

In addition to this, if we compare the $\gamma$ curves performed by LS01 at LEO (sensing a smaller voltage range) with those performed prior to lift-off (see Fig.~\ref{fig:gamma_compara}), we also obtain a result consistent with the change in temperature. This can be observed by the reduction of the slope of the linear part and by the reduction of the maximum $\gamma$ attained, which may be attributed to the increase of the limiting series resistance (R$_+$) as a consequence of its semiconducting-like behavior (see the similarity with Fig.~9 of reference \cite{Acha17}). In other words, no relevant changes on the microscopic factors associated with the transport properties can be determined, other than those linked to the measured temperature variations.

\section{CONCLUSIONS}
A ReRAM device based on two metal-YBCO thin film interfaces was successfully tested over a period of 433 days at LEO by using the LS01 controller. The DUT remains operative after being subjected to the hostile conditions of getting into an orbit entails. Its memory characteristics are evidenced by the existence of RHSLs beside of being subjected to vibrations linked to lift-off, constant thermal cycling ($\pm10~^{\circ}$C, 16x/day) and ionizing radiation representing a cumulative dose 3 orders of magnitude greater than the one it would have received for the same period of time on the surface of the Earth. Gradual changes in the electrical parameters of the ReRAM were observed (R$_{rem}$, V$_{set}$, V$_{reset}$) as the days passed in LEO. However, a detailed study of the electrical transport of the DUT revealed that there were no microscopic changes associated with the existing transport mechanisms. The observed changes could then be explained solely by taking into account the drift of the temperature to which the DUT fixed on the LS01 board was subjected. Incoming studies of the radiation effects at LEO would require the measurement of IV characteristics at constant temperature on denser micrometric arrays of perovskite oxide-metal interfaces, as well as tests over a longer period of time or under higher doses of ionizing radiation. In this way, our results represent a milestone as we move toward a next stage of testing YBCO-based devices. Future radiation dose assessment, without showing signs of malfunction or deviation from their conduction mechanism, would help to prove if perovskite oxide-based ReRAM devices perform better than other memory technologies used in satellites or spacecrafts.

\section{ACKNOWLEDGEMENTS}
We would like to acknowledge financial support by CONICET grants PIP 11220150100322 \& 11220150100653CO, ANPCyT grants PICT 2016-0966 \& 2017-0984 and UBACyT grant {20020170100284BA}
(2018-2020). Jenny and Antti Wihuri Foundation is also acknowledged
for financial support. We are indebted to Satellogic for making possible for us to perform
these experiments on board of their satellites. We thank D.
Gim\'enez, E. P\'erez Wodtke and D. Rodr\'{\i}guez Melgarejo for
their technical assistance.


\end{document}